\documentclass[10pt,aps,prd,twocolumn,nofootinbib,superscriptaddress]{revtex4-2}

\usepackage{natbib}
\usepackage{aas_macros} % aas_macros.sty with a list of journal tag definitions
\usepackage{amssymb,amsmath,accents}
\usepackage{caption} % to spread horizontal alignment in figure captions
\usepackage{subcaption}%,subfig
\usepackage{verbatim}
\usepackage{graphicx}
\usepackage{color,units}
\usepackage[dvipsnames]{xcolor}
\usepackage{soul}
\usepackage{array, makecell, tabularx, cellspace}
\newcolumntype{L}{>{\raggedright\arraybackslash}X}
\newcolumntype{C}{>{\centering\arraybackslash}X}
\usepackage{booktabs} 
\usepackage{bm}
\usepackage{hyperref}
\usepackage{listings}
\usepackage[normalem]{ulem}
\usepackage{amsmath}

\hypersetup{
     unicode=false,          
    pdftoolbar=true,        
    pdfmenubar=true,        
    pdffitwindow=false,     
    pdfstartview={FitH},    
    pdfauthor={Boris Goncharov},     
    colorlinks=true,       
    linkcolor=BrickRed,          
    citecolor=BrickRed,      
    urlcolor=RoyalBlue
}
\graphicspath{{figures/}}

\newcommand{\dd}{\mathrm{d}}
\newcommand{\Msun}{M_{\odot}}

\newcommand{\LamH}{\Lambda_{\mathrm{GSMF}}}

\begin{document}

\definecolor{dkgreen}{rgb}{0,0.6,0}
\definecolor{gray}{rgb}{0.5,0.5,0.5}
\definecolor{mauve}{rgb}{0.58,0,0.82}

\lstset{frame=tb,
  	language=Matlab,
  	aboveskip=3mm,
  	belowskip=3mm,
  	showstringspaces=false,
  	columns=flexible,
  	basicstyle={\small\ttfamily},
  	numbers=none,
  	numberstyle=\tiny\color{gray},
 	keywordstyle=\color{blue},
	commentstyle=\color{dkgreen},
  	stringstyle=\color{mauve},
  	breaklines=true,
  	breakatwhitespace=true
  	tabsize=3
}

\title{Bridging the Population Synthesis of Supermassive Binary Black Holes and the Gravitational Wave Background} 
\author{Kanyuni Iemoto}
 \email{kanyuni.iemoto@utexas.edu}
\affiliation{Max Planck Institute for Gravitational Physics (Albert Einstein Institute), 30167 Hannover, Germany}
\affiliation{Department of Astronomy, University of Texas at Austin, Austin, TX 78712, USA}
\affiliation{Department of Physics, University of Texas at Austin, Austin, TX 78712, USA}
\author{Boris Goncharov}%
\affiliation{Max Planck Institute for Gravitational Physics (Albert Einstein Institute), 30167 Hannover, Germany}
\affiliation{Leibniz Universität Hannover, 30167 Hannover, Germany}
\author{Gabriela Sato-Polito}
\affiliation{School of Natural Sciences, Institute for Advanced Study, Princeton, NJ 08540, United States}
\author{Xiaoming Bi}%
\affiliation{Max Planck Institute for Gravitational Physics (Albert Einstein Institute), 30167 Hannover, Germany}
\affiliation{Leibniz Universität Hannover, 30167 Hannover, Germany}

\date{\today}

\begin{abstract}

Pulsar Timing Arrays (PTAs) constrain population properties of supermassive binary black holes (SMBHBs) through the observation of the gravitational wave background (GWB). 
Unlike other approaches that interpolate population-synthesis libraries or only consider the mean of the strain spectrum, here we capture its full strain probability density directly from semi-analytic population models. 
We apply our new method to the semi-analytic SMBHB population model, independently reproducing the parameter estimation for this model performed by the NANOGrav Collaboration with their 15-yr data.  
We also show the extent to which discrete SMBHB contributions to the GWB resolve degeneracies in the population parameter space.
Finally, using the source-count intensity as the intermediate product in our calculation, we map PTA observations, as a proof of principle, to the SMBHB model based on the galaxy merger prescriptions from numerical hydrodynamical simulations  ``Illustris''. 
We find the effect of delay times $\tau$ between kiloparsec and subparsec SMBHB separations following galaxy mergers, finding $h_{\rm c}$ spanning $(1$--$6)\times10^{-16}$ and $N_{\rm c}$ spanning $(0.3$--$7.1)\times10^{-3}$ for $\tau$ up to $8$~Gyr.

\end{abstract}

\maketitle

\section{Introduction}

A putative gravitational wave background (GWB) in Pulsar Timing Array (PTA) data is expected to contain contributions from a discrete population of supermassive black hole binaries (SMBHBs)\footnote{Possibilities that the GWB is originates from the early universe are also discussed, \textit{e.g.}, in Refs.~\cite{NG15_NEWPHYS,EPTA_DR2_NEWPHYS}. }.
Conventionally, galaxy observations allowed us to indirectly probe masses of galaxies' central black holes by measuring the stellar velocity dispersion $\sigma$ ~\cite{McConnellMa2013,KormendyHo2013}, as well as to estimate galaxy merger rate~\cite{LotzJonsson2011}, which may eventually lead to the formation of SMBHBs~\cite{Sesana2013a,BogdanovicMiller2022}. 
Now, PTAs allow a direct observation of a superposition of a population of SMBHBs at subparsec separations, which explores the final stage of the SMBHB evolution and galaxy mergers. 
Connecting these two observables provides an opportunity to test the physics of how the most massive cosmic objects form and evolve. 

Semi-empirical and semi-analytic population models, as well as cosmological hydrodynamical simulations, aim to connect the evolution of galaxies and their central black holes to the population of SMBHBs observable by PTAs~\cite{JaffeBacker2003}. 
\textit{Astrophysical parameters} $\Lambda$ are quantities that encode aspects of this connection, including the normalization and redshift evolution of the galaxy stellar-mass function and galaxy merger rate, the normalization, slope, and intrinsic scatter of black-hole--host-galaxy scaling relations, the delay between galaxy and black-hole mergers, and the efficiency of dynamical friction, stellar scattering, gas-driven migration, and other binary-hardening processes. 
Semi-empirical models combine observed galaxy demographics with parametrized prescriptions for these processes~\cite{Sesana2013a,MiddletonSesana2021,Sato-PolitoZaldarriaga2024,NG15_HOLODECK}, while semi-analytic galaxy-formation models evolve dark-matter merger trees, galaxies, and black holes using computationally efficient physical prescriptions~\cite{BarausseDvorkin2020,ToubianaSberna2024,CuryloBulik2024}. 
Cosmological hydrodynamical simulations instead coevolve dark matter, gas, stars, and black holes, as in Illustris and Horizon-AGN~\cite{KelleyBlecha2017a,QuelquejayLeclereLi2026}. However, cosmological hydrodynamical simulations do not resolve the subparsec evolution relevant to PTAs.
Their black-hole merger catalogues must therefore be supplemented with subgrid or post-processing prescriptions for orbital decay and binary hardening before PTA observables can be predicted.

There are two limitations of contemporary semi-analytic models with respect to their application to PTA data analysis.
First, they are not generative models.
This means that these models can simulate power spectra of the GWB, but they cannot assess the probability of the power spectra corresponding to certain astrophysical parameters.
Second, their parametrization is too flexible.
This means that many parameters are redundant and the models may have more degrees of freedom than can be constrained by the data.

In particular, the SMBHB population model~\cite{ChenSesana2019} from the recent NANOGrav result~\cite{NG15_HOLODECK} is based on 24 parameters for gravitational-wave-only driven evolution, and an additional 5 parameters for a phenomenological model of SMBHB interactions with the galactic environment such as stars and gas. 
Even though many of these parameters are fixed and only between 4 and 10 are estimated from the PTA data, Ref.~\cite{NG15_HOLODECK} is based on a Gaussian process interpolator between Monte-Carlo strain spectra of the GWB and SMBHB population parameters. 
This introduces an additional layer of approximations which may eventually bias inference. 

The mean characteristic strain can be calculated for many SMBHB population models analytically and fit directly to PTA data. 
A few notable examples are \citet{ChenSesana2019}, \citet{MiddletonSesana2021}, \citet{Sato-PolitoZaldarriaga2024}, \citet{GoncharovSardana2025a}. 
Predictions for PTA observables are also often reduced to mean characteristic strain, \textit{e.g.}, Refs.~\cite{ToubianaSberna2024,Casey-ClydeMingarelli2025}.

Previous analyses did not directly construct the full non-Gaussian distribution $\pi(h_{\rm t}\mid\Lambda)$ which yields realization-dependent characteristic strain spectra $h_{\rm t}(f)$ across gravitational wave frequencies $f$.
For spectral emulation, normalizing flows~\cite{LaalTaylor2025} reproduce simulated strain distributions and their tails more faithfully than Gaussian processes~\cite{NG15_HOLODECK}, although Gaussian processes can remain more accurate for the median spectrum.

In this work, we address the above limitations based on the methodology of \citet{GoncharovSato-Polito2026} and~\citet{Sato-PolitoZaldarriaga2025b}.  
The key principle is that only two effective parameters are sufficient to capture SMBHB properties in the GWB produced by SMBHBs adiabatically inspiralling due to gravitational-wave (GW) emission in circular orbits.
Namely, the mean characteristic strain, $h_{\rm c}$, and the characteristic number of sources, $N_{\rm c}$, at a fiducial reference frequency. 
These parameters can also be recast in terms of the peak of the mass kernel and the number density of SMBHBs. 
They determine the probability density function (PDF) of the total characteristic GW strain, the realizations of which can fluctuate from one frequency to another due to the Poisson process intrinsic to the SMBHB origin of the GWB. 
Refs.~\cite{Sato-PolitoZaldarriaga2024,GoncharovSato-Polito2026} report fitting this model to the 15-year data of the North American Nanohertz Observatory for Gravitational Waves (NANOGrav). 
Here, we apply the same approach to the model from Ref.~\cite{NG15_HOLODECK}.
We further take a step back to calculate the strain PDF in terms of 24 original astrophysical parameters of Ref.~\cite{NG15_HOLODECK}, and fit it to the NANOGrav 15-year data\footnote{We do not directly analyze NANOGrav 15-year data in this work. Instead, we reweigh posteriors to new parametrized priors.}. 
With this, we develop a transparent connection between the 2 primary degrees of freedom that approximate the SMBHB-driven GWB and the astrophysical properties of SMBHBs. 

In addition to the limitations of the semi-analytic models, there are some disagreements in the literature concerning theoretical expectations for the characteristic strain amplitude of the GWB~\cite{Sato-PolitoZaldarriaga2024}. 
Although some discrepancies are attributed to mismodelled noise~\cite{GoncharovSardana2025a}, it remains important to verify the correctness of SMBHB population models and their numerical implementation. 
For example, semi-analytic models in both Ref.~\cite{Sato-PolitoZaldarriaga2025b} and Ref.~\cite{NG15_HOLODECK} are based on the same principle, but use the properties of galaxy velocity dispersion and bulge mass relations, respectively, to map properties of galaxies to those of SMBHBs. 
Thus, systematic errors in SMBHB population models are still under investigation. 
\citet{Simon2023} pointed out that using a velocity-dispersion function (VDF) rather than a galaxy stellar mass function (GSMF) produces a higher $h_{\rm c}$ in the narrower range, with the difference traced partly to additional high-redshift contributions~\cite{Simon2023}.
Underprediction of the galaxy count in the galaxy survey volume may also propagate directly into the local black-hole mass function~\cite{LiepoldMa2024}.
A complementary comparison of velocity dispersion, $K$-band luminosity, and SMBHB mass proxies shows that consistent proxy shifts cancel when the proxy is marginalized, whereas relative calibration offsets between the black-hole and galaxy-count samples can substantially shift the inferred SMBH abundance and predicted GWB amplitude~\cite{Sato-PolitoZaldarriaga2025a}.
These differences are difficult to compare directly because SMBHB population models are often formulated in different astrophysical variables and propagate their assumptions to PTA observables in different ways. Here we place such models on a common methodological foundation by expressing their predictions through the source-count intensity. This provides a common interface for comparing otherwise heterogeneous population models against the same PTA data and facilitates cross-checks of their numerical implementations.

We further demonstrate that our approach can be applied to test the results of numerical hydrodynamic simulations against the GWB observation with PTA data. 
Cosmological hydrodynamical simulations evolve galaxies and SMBHBs with dynamical fractions of gas and feedback. More precisely, they coevolve dark matter and baryons while representing star formation, black-hole accretion, and feedback with subgrid prescriptions, but they do not resolve the subparsec binary evolution that determines residence times in the PTA band.
Hydrodynamic simulations are forward models that predict the SMBHB merger rate. 
Here, we outline the process of calculating the predicted source-count intensity from the hydrodynamic simulations, as well as the consequent pairs of effective parameters, allowing a straightforward comparison with observations.  
We start with galaxy merger rate prescriptions based on the ``Illustris'' simulations \cite{VogelsbergerGenel2013_Illustris_I, GenelVogelsberger2014_illustris_II, TorreyVogelsberger2014_Illutris_III, VogelsbergerGenel2014_Illustris_IV, SijackiVogelsberger2015_Illutris_V} and the population model of \cite{KelleyBlecha2017a}.
We then simulate SMBHB evolution in these merger events from kiloparsec separation to subparsec separation corresponding to the PTA observation band~\cite{KelleyBlecha2017b,ChenDiMatteo2025,NG15_HOLODECK}. 
Thanks to calculating the source-count intensity of these systems, we report the mean characteristic strain and the characteristic number of sources. 

The rest of the paper is organized as follows. 
In Section~\ref{sec:models}, we outline the calculation of the source-count intensity for semi-analytic SMBHB population models. 
In Section~\ref{sec:methods}, we explain the methodology for fitting the SMBHB population model to the NANOGrav 15-year data based on the existing results of Ref.~\cite{GoncharovSato-Polito2026}. 
In Section~\ref{sec:results}, we report the results of parameter estimation. 
In Section~\ref{sec:conclusions}, we present our conclusions. 
    
\section{\label{sec:models}Models} 

A key observable of any SMBHB population model is a \textit{source-count intensity}\footnote{Intensity refers to the strain squared, but we omit this to make terminology less verbose.}~\cite{Sato-PolitoZaldarriaga2024}.
For a frequency bin centered at observed GW frequency $f$ with width $\Delta f$, let $h_s^2$ denote the contribution of one SMBHB source to the squared characteristic strain in that bin, averaged over sky position, inclination, and polarizations. 
The single-source source-count intensity is the rate of change of the number of binaries $N$ with a given $h_s^2$ as a function of $h_s^2$,
\begin{equation}
  \frac{\dd N}{\dd h_s^2}(h_s^2\mid f,\Delta f,\Lambda),
  \label{eq:luminosity-function}
\end{equation}
where $\Lambda$ denotes the parameters of the chosen SMBHB population model. 
An integral over Equation~\ref{eq:luminosity-function} yields the expected number of binaries in that interval.
Its first moment supplies the ensemble strain in the same population-integral sense as the practical theorem of~\citet{Phinney2001}, which relates a cosmological source or remnant population to the GWB spectrum.

We calculate the luminosity function based on Refs.~\cite{NG15_HOLODECK,ChenSesana2019}, which allows us to introduce direct parameter estimation on these models for the first time. 
Ref.~\cite{NG15_HOLODECK} performs Monte-Carlo simulations of the GWB and relies on an interpolator to assemble PDFs from these simulations. 
In contrast, our approach allows us to build a generative model of the GWB from SMBHB following, meaning we both construct a PDF of strain amplitudes of the GWB and perform Monte-Carlo simulations from these PDFs. 

The source-count intensity allows us to apply the same data analysis methodology to two different semi-analytic approaches to SMBHB population properties.
First, to the approach of Refs.~\cite{NG15_HOLODECK,ChenSesana2019} that links the abundance and properties of galaxies to the Galaxy Stellar Mass Function (GSMF), further mapping GSMF to SMBHB properties through the $M_{\rm BH}$-$M_{\rm bulge}$ relation (MMB). 
In particular, \citet{HaringRix2004} measured an $M_{\rm BH}$--$M_{\rm bulge}$ slope of $1.12\pm0.06$ and an observed scatter below $0.30$ dex.
Here, $M_{\rm bulge}$ is the mass of the galactic bulge.
Second, to the approach of Refs.~\cite{Sato-PolitoZaldarriaga2024,GoncharovSato-Polito2026}, where the authors link the abundance and properties of galaxies to their Velocity Dispersion Function (VDF~\citet{BernardiShankar2010}), further mapping VDF to SMBHB properties through the $M_{\rm BH}$-$\sigma$ relation~\cite{GebhardtBender2000,FerrareseMerritt2000}. 
Here, $M_{\rm BH}$ is the SMBHB total mass and $\sigma$ is the galaxy velocity dispersion [km/s].  
The two approaches refer to different galaxy observables, but ultimately represent the same physics. 

\subsection {\label{sec:models:gsmf}Semi-analytic SMBHB population model based on Galaxy Stellar Mass Function}

In this Subsection, we outline the semi-analytic SMBHB population model from Refs.~\cite{NG15_HOLODECK,ChenSesana2019}. 
We perform a calculation in the native galaxy variables
$(\log_{10}m_{\star,1},q_\star,z)$, where $m_{\star,1}$ is the primary galaxy stellar mass, $q_\star=m_{\star,2}/m_{\star,1}\leq1$ is the mass ratio, and $m_{\star,2}$ is the secondary galaxy stellar mass. 
Here, we closely follow Section~3.2.1 of Ref.~\cite{NG15_HOLODECK}.
The GSMF per decade of stellar mass $u \equiv \log_{10}(m_{\star,1}/M_\odot)$ is
\begin{equation}
  \frac{\Psi(\log_{10}m_{\star, 1},z)}{\ln(10)}
  =\,\frac{\Psi_0(z)}{\exp\!\left[\frac{m_{\star, 1}}{M_\Psi(z)}\right]}
   \left(\frac{m_{\star,1}}{M_\Psi(z)}\right)^{1+\alpha_\Psi(z)}
   ,
  \label{eq:gsmf}
\end{equation}
with
\begin{align}
  \log_{10}\Psi_0(z)&=\psi_0+\psi_z z,\\
  \log_{10}\!\left(\frac{M_\Psi(z)}{\Msun}\right)
  &=m_{\psi,0}+m_{\psi,z}z,\\
  \alpha_\Psi(z)&=\alpha_{\psi,0}+\alpha_{\psi,z}z.
\end{align}
The galaxy pair fraction and merger time are power laws,
\begin{align}
  P(m_{\star,1},q_\star,z)
  &=P_0\left(\frac{m_{\star,1}}{10^{11}\Msun}\right)^{\alpha_p}
      (1+z)^{\beta_p}q_\star^{\gamma_p},\\
  T_{\rm gal}(m_{\star,1},q_\star,z)
  &=T_0\left(\frac{m_{\star,1}}{M_{t,0}}\right)^{\alpha_t}
      (1+z)^{\beta_t}q_\star^{\gamma_t}.
\end{align}
Following Ref.~\cite{NG15_HOLODECK}, let us distinguish the galaxy-pair redshift \(z_g\) from the post-delay SMBHB redshift \(z_f\).
The galaxy-merger number density is therefore
\begin{equation}
\frac{\dd^3n_{\rm gal}}
     {\dd u\,\dd q_\star\,\dd z_g}
=
\Psi(m_{\star,1},z_g)
\frac{P(m_{\star,1},q_\star,z_g)}
     {T_{\rm gal}(m_{\star,1},q_\star,z_g)}
\left|\frac{\dd t}{\dd z_g}\right|.
  \label{eq:galaxy-merger-measure}
\end{equation}
The black-hole--bulge relation applied to each galaxy is
\begin{equation}
  \log_{10}\!\left(\frac{M_{\rm BH}}{\Msun}\right)
  =\mu+\alpha_\mu\log_{10}\!\left(
    \frac{f_{\rm bulge}m_\star}{10^{11}\Msun}\right)
   +\mathcal{N}(0,\epsilon_\mu).
  \label{eq:mmb}
\end{equation}
The galaxy-pair density is mapped to the two black-hole component-mass axes, and the numerical scatter redistribution of the imported ``GW-only'' model of \textsc{holodeck} is applied to that component-mass grid.
This approach follows Section~3.2.2 of Ref.~\cite{NG15_HOLODECK}.

The GSMF approach is therefore based on a total of 24 parameters. 
Following the case of EM-agnostic GW-only SMBHB inspirals in Ref.~\cite{NG15_HOLODECK}, we perform parameter estimation on only four parameters 
\begin{equation}
  \LamH=\{\psi_0,m_{\psi,0},\mu,\epsilon_\mu\},
\end{equation}
and we fix the remaining parameters to the values specified in Table~B2 in Ref.~\cite{NG15_HOLODECK}.
For the astrophysically informed analysis, we adopt the priors from Ref.~\cite{NG15_HOLODECK}: $\pi(\psi_0) = \mathcal{N}(-2.56,0.4)$, $\pi(m_{\psi,0})=\mathcal{N}(10.9,0.4)$, $\pi(\mu) = \mathcal{N}(8.6,0.2)$, and $\pi(\epsilon_\mu)=\mathcal{N}(0.32,0.15)\,\mathrm{dex}$.
The priors on the GSMF normalization and characteristic mass were obtained from fits to the measurements of \citet{TomczakQuadri2014}.

\subsection{\label{sec:models:dndhs2}Source-count intensity}

In this Subsection, we describe the source-count intensity for the semi-analytic SMBHB population parametrization outlined in Subsection~\ref{sec:models:gsmf}. 
Following the prescription from Ref.~\cite{Sato-PolitoZaldarriaga2024}, it is calculated as the integral over a decade of total mass, mass ratio, and redshift:
\begin{equation}
    \begin{split}
        \frac{\dd N}{\dd \log_{10}h_s^2} 
        &= \int \dd \log_{10} M \int \dd q \int \dd z \ \delta 
        \left(\log_{10}h_s^2 - \log_{10}\tilde{h}_s^2
        \right) \\
        &\quad\times
        \frac{\dd N}{\dd \log_{10}M \ \dd q  \ \dd z  \ \dd lnf}.
    \end{split}
    \label{eq:dNdlog10hs2_delta}
\end{equation}
Here, $\tilde{h_s^2}$ is a function of the total mass $M$ and comoving radial distance $\chi(z)$
\begin{equation}
    \tilde{h_s^2} 
    =  \frac{32}{5} 
    \frac{(GM)^{10/3}}{c^8} \eta^2  \frac{(\pi f[1+z])^{4/3}}{\chi^2(z)} \frac{f}{\Delta f},
    \label{eq:tilde_h2s}
\end{equation}
so that the integral over decade total mass in Equation~\ref{eq:dNdlog10hs2_delta} is calculated by changing the integration variable mass to strain, introducing the Jacobian, and solving the delta function in $h_{\rm s}$, assuming a finite frequency resolution of $\Delta \ln f \approx  \Delta f / f $ and $\Delta f = T_\mathrm{{obs}}^{-1}$. 
Effectively, $\tilde{h_s^ 2}$ is an argument of the source-count intensity and $h_s^2$ is the value predicted from population coordinates.
The total number of SMBHBs per logarithmic frequency and binary parameters is calculated as a by-product of the calculations in Ref.~\cite{NG15_HOLODECK}.
\begin{equation}
    \frac{\dd^4 N}{\dd\log_{10}M\,\dd q\,\dd z\,\dd\ln f}
    =\frac{\dd n}{\dd z \dd \log_{10} M \dd q}
     \frac{\dd t_r}{\dd\ln f_r}
     \left|\frac{\dd z}{\dd t_r}\right|
     \frac{\dd V_c}{\dd z}.
    \label{eq:dNdp}
\end{equation}
Term $\frac{\dd n}{\dd \log M \ \dd q \ \dd z} $ is calculated based on Equation~\ref{eq:galaxy-merger-measure} and Equation~\ref{eq:mmb}.

From the source-count intensity, we calculate the two observational degrees of freedom of the GWB from SMBHBs as
\begin{align}
    h_{\mathrm{s,peak}}^2
    &= \arg\max_{h_s^2}\!\left[h_s^2\,\frac{dN}{d\log h_s^2}\right],
    \label{eq:h2peak-def}\\
    h^2_c
    &= \displaystyle \int h_s^2 \frac{dN}{d \log h_s^2}\,d \log h_s^2.
    \label{eq:Nc-def}
\end{align}
The characteristic number of sources is defined as 
\begin{equation}
    N_c \equiv \frac{h_c^2}{ h_{\mathrm{s,peak}}^2}.
    \label{eq:hc2-def}
\end{equation}

\begin{figure*}
    \centering
    \includegraphics[width=\textwidth]{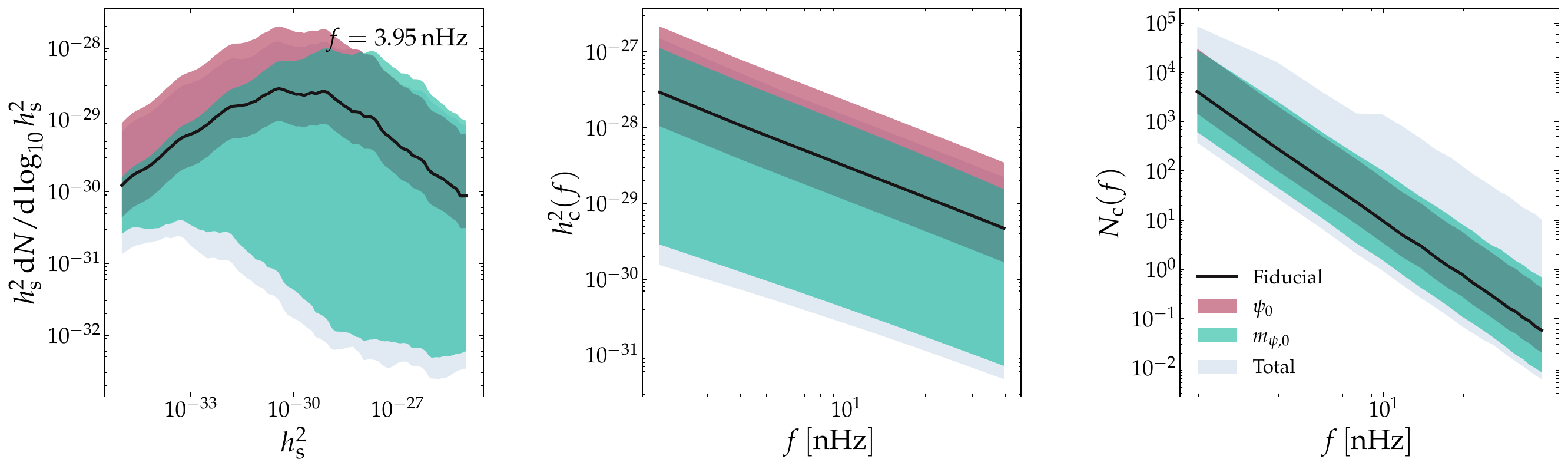}
    \caption{The impact of SMBHB population properties on the source-count intensity at a fiducial frequency of $f = 4~\rm{nHz}$ (left panel) and two PTA observables, $(h_c, N_c)$, shown as a function of frequency (middle panel and left panel, respectively). 
    The shaded area covers a $3\sigma$ area of Monte-Carlo draws for population hyperparameters $\Lambda_{\rm GSMF}$ from the Normal ``Astrophysical'' priors specified in Table B1 in Ref.~\cite{NG15_HOLODECK}. 
    The solid line corresponds to the central values of $\Lambda_{\rm GSMF}$ in these priors. 
    The dashed lines correspond to the central values for all parameters except a single parameter, for which the 5th percentile value is chosen. 
    Similarly, dash-dotted lines correspond to the 95th percentile. 
    Red shades correspond to the parameter $\psi_0$ and green lines correspond to the present-time galactic mass parameter $m_{\psi,0}$. 
    } 
    \label{fig:dndhs2_hc_Nc}
\end{figure*}

We demonstrate the impact of SMBHB properties on the source-count intensity as well as on the PTA degrees of freedom, $h_c$ and $N_c$, in Figure~\ref{fig:dndhs2_hc_Nc}. 
The left panel shows the luminosity function times $h^2_s$, which is the argument of $\arg \max$ in Equation~\ref{eq:h2peak-def}. 
Thus, the peak position corresponds to $h^2_{\rm s,peak}$. 
The function is shown for a range of astrophysical parameters at a fiducial frequency of $f = 4~\rm{nHz}$. In addition to showing the order of magnitude span of this function for astrophysically motivated $\Lambda_{\rm GSMF}$, the panel shows the impact of SMBHB abundance and mass scale. 
The abundance parameter $\psi_0$ only changes the normalization of the source-count intensity, moving the curve vertically but not horizontally. 
The mass scale, however, affects both the normalization and the position of $h^2_{\rm s,peak}$. 
It is the result of the relation of $h^2_{\rm s,peak}$ to the peak of the mass kernel~\cite{Sato-PolitoZaldarriaga2025b}, meaning that reducing the mass of SMBHBs reduces the strain at which they contribute the most. 
The right two panels demonstrate that increasing either the SMBHB abundance or the mass scale increases both $h_c$ and $N_c$. 

\subsection{Prescriptions from the cosmological simulation ``Illustris''}
\label{sec:illustris-prescriptions}

In addition to considering semi-analytic models, we demonstrate the calculation of the source-count intensity for numerical cosmological galaxy simulations. 
Illustris is a suite of cosmological hydrodynamical simulations that follows the coupled evolution of dark matter, gas, stars, and massive black holes with the moving-mesh code \textsc{Arepo}~\cite{VogelsbergerGenel2013_Illustris_I,VogelsbergerGenel2014_Illustris_IV,GenelVogelsberger2014_illustris_II}.
The highest-resolution realization, Illustris-1, evolves a periodic comoving volume $(106.5~\mathrm{Mpc})^3$ from $z=127$ to $z=0$, with baryonic mass resolution $1.26\times10^6\,M_\odot$ and a minimum gravitational softening scale of approximately $710$~pc at $z=0$~\cite{VogelsbergerGenel2014_Illustris_IV, NelsonPillepich2015}.
Its galaxy-formation model includes radiative cooling, star formation and stellar evolution, chemical enrichment, stellar feedback, and prescriptions for massive-black-hole seeding, gas accretion, black-hole feedback, and mergers~\cite{VogelsbergerGenel2013_Illustris_I,SijackiVogelsberger2015_Illutris_V}.

Massive black holes are represented by sink particles whose masses and accretion histories are evolved together with their host galaxies.
When two black-hole particles approach within a gravitational smoothing length, typically of order a kiloparsec, Illustris records a black-hole ``merger''~\cite{KelleyBlecha2017a,KelleyBlecha2017b}.
This event should not be identified with the physical coalescence of a bound SMBHB: the simulation does not resolve the subsequent orbital decay through parsec and subparsec separations.
The merger catalogue constructed from Illustris associates the component black-hole masses and merger epoch with properties of the two host subhalos, including their stellar masses, stellar half-mass radii, and velocity dispersions.

Predicting a nanohertz GW signal therefore requires a post-processing prescription that evolves each catalogue system from galactic scales to coalescence.
In the present work we retain the Illustris prediction for the black-hole population and merger epochs but replace those detailed environmental prescriptions with the phenomenological hardening model described in Section~\ref{sec:methods}.

\section{\label{sec:methods}Methodology}

\subsection{Bayesian inference}

One of the aims of this work is to extend parameter estimation by~\citet{GoncharovSato-Polito2026} on the NANOGrav 15-year data~\cite{NG15_data} directly to a set of astrophysical GSMF parameters. 
The posterior is expressed as 
\begin{equation}
    \mathcal{P}(\theta,h_{t},\Lambda|t_a) \propto \mathcal{L}(t_a|h_t,\theta) \pi(h_t|\Lambda) \pi(\Lambda) \pi(\theta),
    \label{eq:posterior}
\end{equation}
where $\mathcal{L}$ is the PTA likelihood marginalized over reduced-rank coefficients~\cite{NG9_GWB}, $\pi$ is the parametrized prior, $h_t$ are characteristic strain amplitudes of the GWB, $\theta$ are the remaining PTA noise parameters, $\Lambda$ are hyperparameters, $\pi$ is the hyperprior. 
In Ref.~\cite{GoncharovSato-Polito2026}, $\Lambda=(N_c,h_c)$.
The initial source-count intensity is obtained based on the VDF approach discussed at the end of Subsection~\ref{sec:models:dndhs2} and rescaled for subsequent $(N_c,h_c)$ samples~\cite{Sato-PolitoZaldarriaga2025b}. 
Here, we perform posterior reweighting of the results of Ref.~\cite{GoncharovSato-Polito2026} to $\Lambda=\LamH$.
Based on the source number count in Equation~\ref{eq:luminosity-function}, we build $\pi(h_{{\rm t},i} \mid \LamH)$ directly, following the methodology of Refs.~\cite{Sato-PolitoZaldarriaga2024,GoncharovSato-Polito2026}.
Thus, using the terminology of importance sampling, $\pi(h_{{\rm t},i} \mid N_c,h_c)$ may be referred to as the proposal component, and $\pi(h_{{\rm t},i} \mid \LamH)$ is referred to as the target component. 

Examining the right-hand side of Equation~\ref{eq:posterior} applied to the target distribution, expanding unity in terms of the ratio of the target distribution to itself, and rearranging the terms, we obtain
\begin{equation}
\begin{split}
    \mathcal{P}(\theta,h_{t},\LamH|t_a) \propto \mathcal{P}(\theta,h_{t},N_c,h_c|t_a) \times \\ \times \frac{\pi(h_t|\LamH)}{\pi(h_t|N_c,h_c)} \frac{\pi(\LamH)}{\pi(N_c,h_c)}.
\end{split}
\end{equation}
Marginalizing the posterior over $(\theta,h_t,N_c,h_c)$, and using the expression for Monte-Carlo integration 
\begin{equation}
    \langle f(x) \rangle_{p(x)} = \int f(x) p(x) dx \approx \frac{1}{n_\text{s}} \sum_{i}^{n_\text{s}}f(x_i),
    \label{eq:integralsum}
\end{equation}
we obtain our expression for the target posterior
\begin{equation}
    \mathcal{P}(\LamH|t_a) \approx \frac{1}{n_s}\sum^{n_s}_i \frac{\pi(h_{t,i}|\LamH)}{\pi(h_{t,i}|N_{c,i},h_{c,i})} \frac{\pi(\LamH)}{\pi(N_{c,i},h_{c,i})}.
    \label{eq:posterior_reweighted}
\end{equation}
The elements of the sum in the equation above are referred to as weights, $w_i(\LamH)$.
The posterior is sampled using rejection sampling. 
For each proposal indexed $j$, we draw $u_j\sim\mathcal{U}(0,1)$ and accept the point when
\begin{equation}
  u_j<\frac{w_j}{w_{\max}},
  \qquad
  w_{\max}=\max_j w_j.
  \label{eq:rejection}
\end{equation}
Rejection sampling is a Monte Carlo algorithm: proposals are drawn from a known density and retained with probability proportional to the ratio of the target density to the proposal density.
Equation~\ref{eq:rejection} uses the largest weight in the finite proposal set as the envelope.
The posterior we obtain using this method is therefore a rejection-resampled approximation to Equation~\ref{eq:posterior_reweighted}.
It is an exact rejection sample only if $w_{\max}$ also bounds the importance ratio over the full support of the new hyperprior.
For validity, the proposal must have support wherever the target posterior has appreciable mass.
Otherwise, no finite set of weights can reconstruct the missing region.
A practical overlap diagnostic is the effective sample size $n_{\rm eff}=(\sum_i w_i)^2/\sum_i w_i^2$~\cite{HourihaneMeyers2023}.
A small value indicates that a few proposals dominate the importance estimate.
Normalized weighted sums are the direct importance-sampling estimates, while Equation~\ref{eq:rejection} produces a convenient finite, unweighted approximation.

\subsection{Estimation of GWB with Illustris}

We use the discrete Illustris population distributed with \textsc{holodeck}, which was assembled from the black-hole and host-subhalo catalogue of Refs.~\cite{KelleyBlecha2017b,KelleyBlecha2017a}.
Catalogue entry $j$ supplies the two black-hole masses, the scale factor at the recorded Illustris merger event, and the stellar half-mass radii of the two hosts.
Following Section~\ref{sec:illustris-prescriptions}, we initialize each system at a separation $a_{j,\mathrm{init}}$ equal to the sum of these two stellar half-mass radii.

We assume circular binaries and evolve their separation with the phenomenological two-power-law hardening rate used by \textsc{holodeck} \cite{KelleyBlecha2017b, NG15_SMBHB}, 
\begin{equation}
 \frac{\dd a}{\dd t_r}
 =-A_j\frac{(1+x)^{-\gamma_{\rm out}+\gamma_{\rm in}}}{x^{\gamma_{\rm in}-1}}
 +\left.\frac{\dd a}{\dd t_r}\right|_{\rm GW},
 \qquad x\equiv\frac{a}{a_c},
 \label{eq:illustris-hardening}
\end{equation}
where the first term represents environmental hardening and the second is the quadrupolar GW backreaction for a circular binary.
We adopt the implementation defaults $a_c=100~\mathrm{pc}$, $\gamma_{\rm in}=-1$, and $\gamma_{\rm out}=1.5$.
For every catalogue binary, the normalization $A_j$ is chosen so that the evolution from $a_{j,\mathrm{init}}$ to coalescence has a specified total lifetime $\tau$.
Thus, $\tau$ is a parameter that summarizes the unresolved orbital decay after the catalogue event.
The binary scale factor is evolved simultaneously, so only systems that reach a selected observer-frame frequency before $z=0$ contribute at that frequency.

As stated in \cite{KelleyBlecha2017a} for each observed GW frequency $f_i$, only the catalogue binaries indexed $j$ that reach this frequency before $z=0$, and their numerically evolved orbital trajectories are interpolated to the corresponding rest-frame orbital frequency$f_{r,ij}=(1+z_{ij})f_i/2$.
We then calculate its sky- and polarization-averaged source strain $h_{s,ij}^2$\footnote{Code \textsc{holodeck} returns the orientation-averaged root-mean-squared (RMS) time-domain strain, $h_{s}=h_{\rm rms}=\sqrt{8/5}\,h_0$, rather than characteristic strain. Following the strain conventions reviewed in Appendix~A of Ref.~\cite{GoncharovSato-Polito2026}, its contribution in one PTA Fourier bin is $\widetilde h_s^2=h_{\rm rms}^2/\Delta\ln f\simeq h_{\rm rms}^2f/\Delta f$.}.
For the Illustris-1 catalogue, we use the physical comoving volume $V_{\rm sim}=(75/h)^3~\mathrm{Mpc}^3=(106.5~\mathrm{Mpc})^3$, with $h=0.704$. 
Following \cite{KelleyBlecha2017a, ChenDiMatteo2025}, single catalogue entry represents a comoving density $V_{\rm sim}^{-1}$, and its expected number of light-cone sources per logarithmic frequency is
\begin{equation}
 W_{ij}
 =\frac{1}{V_{\rm sim}}
 \frac{\dd V_c}{\dd z}\left|\frac{\dd z}{\dd t_r}\right|
 \frac{\dd t_r}{\dd\ln f_r}
 =\frac{4\pi c(1+z_{ij})\chi^2(z_{ij})}{V_{\rm sim}}
 \frac{f_{r,ij}}{\dot f_{r,ij}}.
 \label{eq:illustris-weight}
\end{equation}
Here $\chi(z)$ is the radial comoving distance. The expected multiplicity of this entry in a finite observed-frequency bin is $\lambda_{ij}=W_{ij}\Delta\ln f_i$, and a realization draws the corresponding integer multiplicity from $\operatorname{Poisson}(\lambda_{ij})$.
Consequently, no continuous mass-to-strain Jacobian is required: the source-count intensity is a weighted empirical distribution over the evolved catalogue.
For a bin $k$ in $\log_{10}\widetilde h_s^2$, it is
\begin{equation}
 \left.\frac{\dd N_i}{\dd\log_{10}\widetilde h_s^2}\right|_k
 \simeq
 \frac{1}{\Delta\log_{10}\widetilde h_{s,k}^2}
 \sum_j\lambda_{ij} \Bigg|_k \,,
 \qquad
 \widetilde h_{s,ij}^2\equiv\frac{h_{s,ij}^2}{\Delta\ln f_i},
 \label{eq:illustris-weighted-histogram}
\end{equation}
where we select catalogue entries whose strain lies in bin $k$, and the factor $(\Delta\ln f_i)^{-1}$ converts the instantaneous source strain to its contribution to characteristic strain squared in the Fourier bin~\cite{Sato-PolitoZaldarriaga2025b}.
The first moment is equivalently
\begin{equation}
 \left\langle h_c^2(f_i)\right\rangle
 =\sum_j\lambda_{ij}\widetilde h_{s,ij}^2
 =\sum_j W_{ij}h_{s,ij}^2,
 \label{eq:illustris-hc2-consistent}
\end{equation}
which explains why the bin-width factors cancel in the mean characteristic strain.
We obtain $\widetilde h_{s,\mathrm{peak}}^2$ from the maximum of $\widetilde h_s^2\dd N_i/\dd\log_{10}\widetilde h_s^2$ and define $N_c=\langle h_c^2\rangle/\widetilde h_{s,\mathrm{peak}}^2$, using the same Fourier-bin convention as the PTA posterior.

We estimate the internal finite-catalogue uncertainty with a spatial block bootstrap at 12 fixed values of \(\tau\) between \(0.01\) and \(8~\mathrm{Gyr}\).
For each \(\tau\), the calculation proceeds in three steps.
First, we evolve the original Illustris catalogue and calculate the strain and light-cone contribution of every binary reaching \(f_{\rm ref}=1\,\mathrm{yr}^{-1}\).
Second, we divide the simulation volume into \(3^3=27\) equal cubic blocks and assign each merger event to a block using the midpoint of its two progenitor-subhalo positions along their shortest periodic separation.
Third, we draw 27 blocks with replacement and multiply the contribution of every binary by the number of times its block was drawn.
For example, binaries in a block drawn twice contribute twice, while binaries in an undrawn block do not contribute to that realization.
We repeat this block draw 5000 times and, for each resulting reweighted catalogue, recalculate \(\langle h_c^2\rangle\), the strain-kernel peak \(\widetilde h_{s,\mathrm{peak}}^2\), and their ratio \(N_c\).
Because the maximum of the raw 250-bin strain kernel can be set by a single high-strain catalogue entry, we estimate its mode after Gaussian-kernel interpolation in $\log_{10}h_s^2$ with bandwidth $0.25$ dex.
Changing this bandwidth from $0.15$ to $0.35$ dex changes the central $\log_{10}N_c$ by at most $0.08$ dex over the sampled lifetimes.
This procedure propagates spatial finite-catalogue fluctuations in the simulation into both plotted coordinates without adding physical-sky Poisson draws, whose source discreteness is already described by the likelihood.
It cannot sample density modes larger than the Illustris box, rare massive binaries absent from the catalogue, or independent cosmological initial conditions, and is therefore an internal uncertainty estimate rather than a complete cosmic-variance interval.

\begin{figure}
    \centering
    \includegraphics[width=1.0\linewidth]{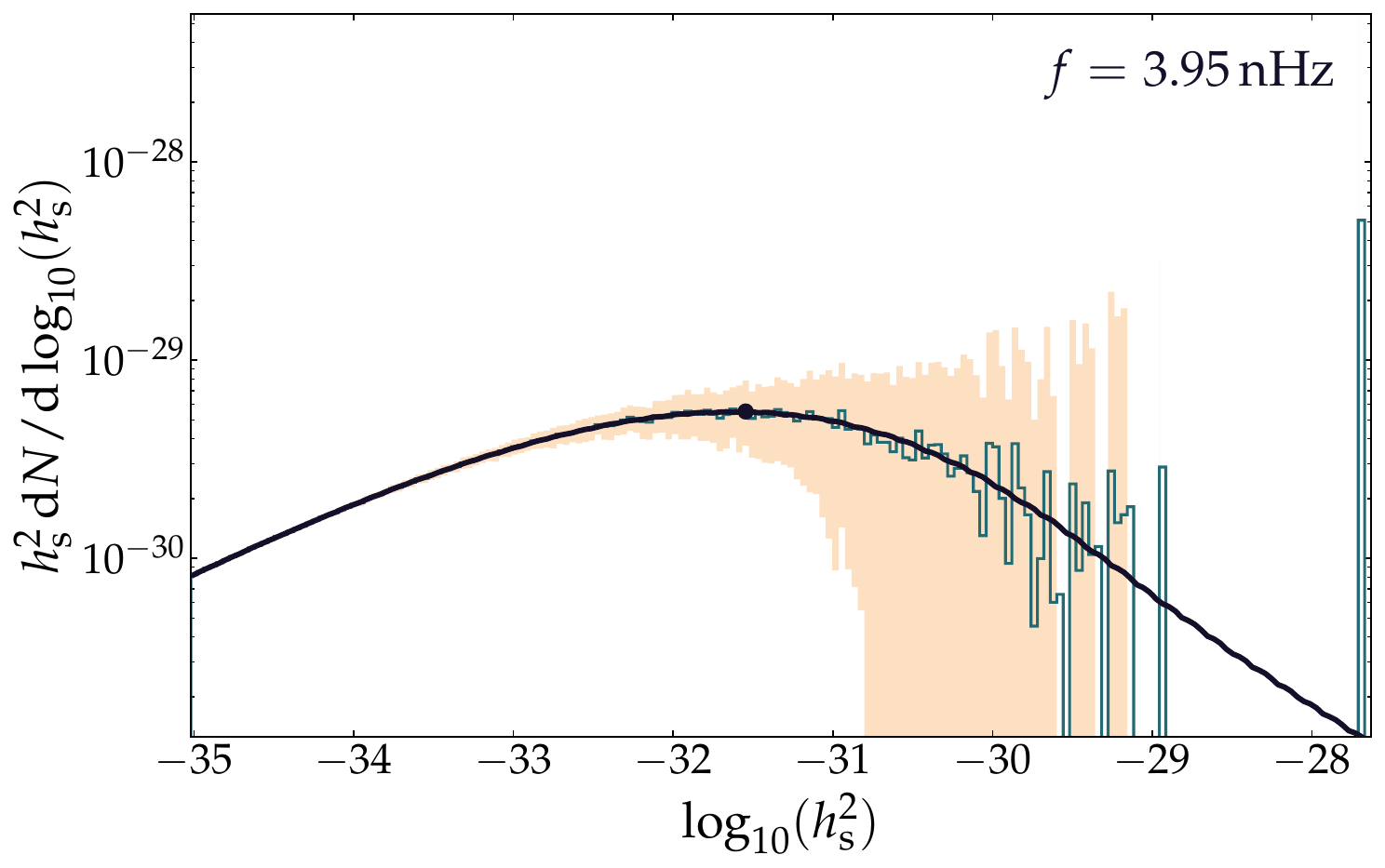}
    \caption{Validation of the numerical source-count intensity calculation using the fiducial semi-analytic population.  The solid black curve is the analytical source-count intensity kernel at $4$~nHz, the blue stepped curve is its numerical counterpart, and the shaded orange region shows the numerical mean and $1\sigma$ realization-to-realization scatter obtained from 100 Monte Carlo realizations. 
    A circle marks the maximum.  
    The numerical calculation agrees with the analytical calculation up to the high-strain tail dominated by cosmic variance.}
    \label{fig:dndhs2_numerical}
\end{figure}

Before applying this numerical construction to Illustris, for which no closed-form source-count intensity is available, we validate it using the fiducial semi-analytic population $\Lambda_{\rm GSMF}$.  For this control population, the source-count intensity can be evaluated both analytically and from Monte Carlo realizations. In Figure~\ref{fig:dndhs2_numerical}, we test the construction of the strain density kernel used to identify $(N_{\rm c},h_{s,\mathrm{peak}}^2)$.
The widening scatter above the maximum arises because the expected occupancy of a high-strain bin is small, so the presence or absence of rare, loud binaries produces large Poisson fluctuations~\cite{Sato-PolitoZaldarriaga2025b}. 
Analogous source discreteness is expected for Illustris.  Illustris additionally supplies only a finite cosmological volume and merger catalogue, so its estimated strain kernel can itself fluctuate: if those fluctuations shift the kernel maximum or cause rare systems to dominate its first moment, they propagate jointly to $h_{s,\mathrm{peak}}^2$ and $N_c$.  The spatial block-resampling calculation described above estimates this latter, internal finite-catalogue contribution.

\section{\label{sec:results}Results}

\subsection{Illustris}

\begin{figure}
    \centering
    \includegraphics[width=1\linewidth]{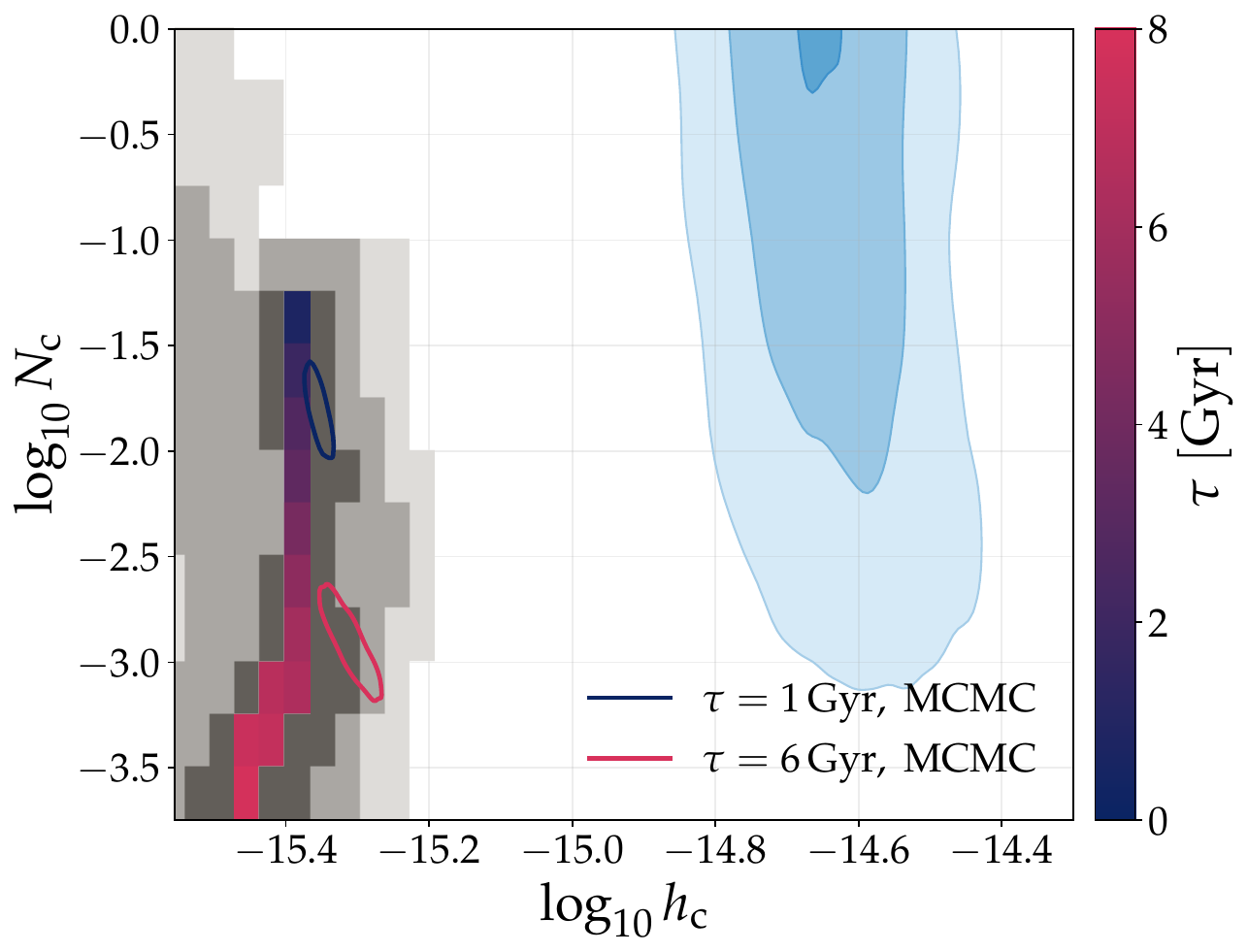}
    \caption{A prediction of PTA observables $(N_c,h_c)$ from the cosmological ``Illustris'' simulation, compared against the results of parameter estimation based on the NANOGrav 15-year data~\cite{GoncharovSato-Polito2026}. 
    The color of pixels corresponds to a delay time between galaxy mergers and SMBHB mergers modeled according to Ref.~
    \cite{NG15_HOLODECK}.
    Gray pixels show the union of conditional, fixed-$\tau$ internal finite-catalogue regions from spatial block resampling, with darker shades denoting the two-dimensional Gaussian-equivalent $1\sigma$, $2\sigma$, and $3\sigma$ levels.
    Hollow contours mark the two-dimensional Gaussian-equivalent $1\sigma$ contours from fits to two individual Illustris realizations.
    }
    \label{fig:illustris}
\end{figure}

In Figure~\ref{fig:illustris}, we calculate Illustris predictions for $(N_{\rm c},h_{\rm c})$ at the reference frequency $f_{\rm ref}=1\,\mathrm{yr}^{-1}$ shown as colored and grey cells. 
For colored cells, Illustris catalogue and hardening-shape parameters are held fixed while the source-count intensity is evaluated at $10^5$ lifetimes drawn uniformly in $\tau\in[0.01,8]~\mathrm{Gyr}$. 
Thus, the samples are not independent SMBHB population realizations. 
The colored cells show the mean $\tau$ of the deterministic model evaluations falling in each occupied cell of the $(\log_{10}h_c,\log_{10}N_c)$ plane.
The cells with three shades of grey show the union, over the 12 fixed lifetimes, of the two-dimensional Gaussian-equivalent $1\sigma$, $2\sigma$, and $3\sigma$ regions obtained by spatial block resampling.
Grey cells approximately correspond to the effect of drawing Monte-Carlo realizations of $h_{\rm t}$ of the GWB. 
The hollow dark-blue and pink-red contours show the two-dimensional Gaussian-equivalent $1\sigma$ regions obtained by fitting individual Illustris realizations at $\tau=1$ and $6~\mathrm{Gyr}$, respectively, using Markov-Chain Monte Carlo (MCMC).
For each fit, the 20 per-frequency values of $h_t$ from one Illustris realization are held fixed, $\pi(h_t\mid N_c,h_c)$ is evaluated as the likelihood, and the priors $\pi(N_c)$ and $\pi(h_c)$ are uniform. 
Unlike the source-count intensity, which is our new calculation based on \textsc{holodeck}, MCMC contours are calculated on the direct output from \textsc{holodeck} population synthesis. 
To compare the simulations with observations, the blue contours enclose the posterior $1$, $2$, and $3\sigma$ credible levels from the analysis of the NANOGrav 15-yr data~\cite{GoncharovSato-Polito2026}. 
We only show a segment of the posterior which is near our Illustris prediction; the remaining segment is irrelevant to the discussion. 

All Illustris results quoted here use the physical simulation volume $V_{\rm sim}=(75/h)^3\,\mathrm{Mpc}^3$ with $h=0.704$, and bin-dependent quantities use $\Delta f=T_{\rm obs}^{-1}$ with $T_{\rm obs}=16.03\,\mathrm{yr}$.
Across the explored lifetimes, Illustris predicts $h_c(1\,\mathrm{yr}^{-1})=(3.40$--$4.27)\times10^{-16}$.
The NANOGrav posterior samples used in the comparison instead have median $h_c=2.17\times10^{-15}$ and a 68\% interval $(1.85$--$2.57)\times10^{-15}$.
The Illustris prediction therefore lies below the displayed NANOGrav posterior contours for every lifetime considered, with a characteristic-strain amplitude approximately a factor of five to six below the posterior median.
With this convention, the conditional Illustris distribution has median $\log_{10}N_c=-2.48$ and 95\% interval $[-3.57,-2.15]$ at $f_{\rm ref}=1\,\mathrm{yr}^{-1}$.
This lower $N_c$ corresponds to a smaller characteristic number of contributors and hence stronger source discreteness, but it remains conditional on the sampled lifetime distribution, finite-volume catalogue, and other fixed population assumptions.

The outer $3\sigma$ internal finite-catalogue region for Illustris also remains disjoint from the NANOGrav posterior contours.
Varying $\tau$ alone does not reconcile the fixed Illustris SMBHB population with the measured GWB amplitude.
This is not unexpected: the literature compilation in Ref.~\cite{NG15_HOLODECK} gives $A_{\rm yr}=(1$--$6)\times10^{-16}$ for the Illustris-based model suite of Ref.~\cite{KelleyBlecha2017a}, overlapping with the range found here. 
The amplitude difference between the NANOGrav posterior and the Illustris prediction is a factor of five to six in characteristic strain.
It reflects discrepancies in the abundance or masses of SMBHBs with dominant contributions, as well as the fractions of SMBHBs that reach the PTA band.
Alternative binary evolution prescriptions, the increased volume of the simulation, as well as modifiers to the catalogued masses are the factors that may improve the consistency of our Illustris prediction and the posterior. 

Let us examine the impact of $\tau$ on the results more closely. 
Long lifetimes generally move the Illustris prediction toward lower $h_c$ and lower source counts because fewer catalogue binaries reach the PTA band by the present epoch.
The detailed track is not strictly monotonic because changing the lifetime changes which members of the finite Illustris catalogue reach the reference frequency and which strain bin sets $h_{s,\mathrm{peak}}^2$.
The comparatively weak dependence of $h_c$ on $\tau$ over much of the sampled interval can be understood from the residence-time weighting.
At representative lifetimes spanning the plotted range, more than 99\% of the calculated $h_c^2$ at $f_{\rm ref}$ is contributed by binaries for which GW emission supplies at least 90\% of the instantaneous hardening rate.
Specifically, multiplying the finite-bin source contribution in Equation~\ref{eq:tilde_h2s} by its expected multiplicity $\lambda=W\Delta\ln f$ cancels the bin-width factors, as in Equation~\ref{eq:illustris-hc2-consistent}.
Combining the remaining single-source strain with the light-cone weight in Equation~\ref{eq:illustris-weight}, and using the GW-driven relation $\dot f_r\propto\mathcal{M}^{5/3}f_r^{11/3}$, gives $h_s^2W\propto\mathcal{M}^{5/3}(1+z)^{-1/3}f^{-4/3}$.
In this product, the inverse-square decrease of an individual source’s strain, \(h_s^2\propto \chi^{-2}\), is offset by the \(\chi^2\) factor in the differential comoving volume, \(\mathrm{d}V_c/\mathrm{d}z\propto \chi^2/H(z)\), which counts sources in a shell at that redshift.
Consequently, changing $\tau$ while retaining nearly the same set of catalogue mergers changes their contribution only through the weak factor $(1+z)^{-1/3}$, explaining the near-plateau at short lifetimes.
For the shortest-lifetime realization, the lower tenth percentile of the $h_c^2$-weighted catalogue lookback times is approximately $3.9~\mathrm{Gyr}$. 
Once $\tau$ becomes comparable to this scale, GWB-relevant late mergers begin to fail to reach the PTA band by $z=0$, explaining why the amplitude becomes more sensitive to $\tau$ only at several-Gyr lifetimes.
At large $\tau$, the dominant change in the population occurs when another member of the finite catalogue crosses this boundary, so locally flat or stepped portions of the track should be regarded as catalogue discreteness.

There is something to keep in mind when considering $\tau<1~{\rm Gyr}$.
This GW dominance at $f_{\rm ref}$ does not extend over the entire PTA band for the shortest lifetimes: the calculated Illustris spectra show a pronounced low-frequency turnover for $\tau=0.01~\mathrm{Gyr}$, a weaker departure confined to the lowest frequency bin for $\tau\simeq0.08$--$0.5~\mathrm{Gyr}$, and no comparable low-frequency suppression larger than approximately 10\% for $\tau\simeq1$--$8~\mathrm{Gyr}$.
The turnover occurs because enforcing a shorter total lifetime requires stronger phenomenological environmental hardening, which reduces the binary residence time at low frequencies; in the $\tau=0.01~\mathrm{Gyr}$ model, the suppression remains visible through $f\simeq4.8~\mathrm{nHz}$ but is negligible by $f_{\rm ref}$.
It therefore explains the low-frequency spectral shape of the short-$\tau$ models, but not the near-plateau of $h_c(1\,\mathrm{yr}^{-1})$ with $\tau$.

\subsection{Insights from the NANOGrav 15-year data}

\begin{figure*}[!htb]
  \centering
  \includegraphics[width=\textwidth]{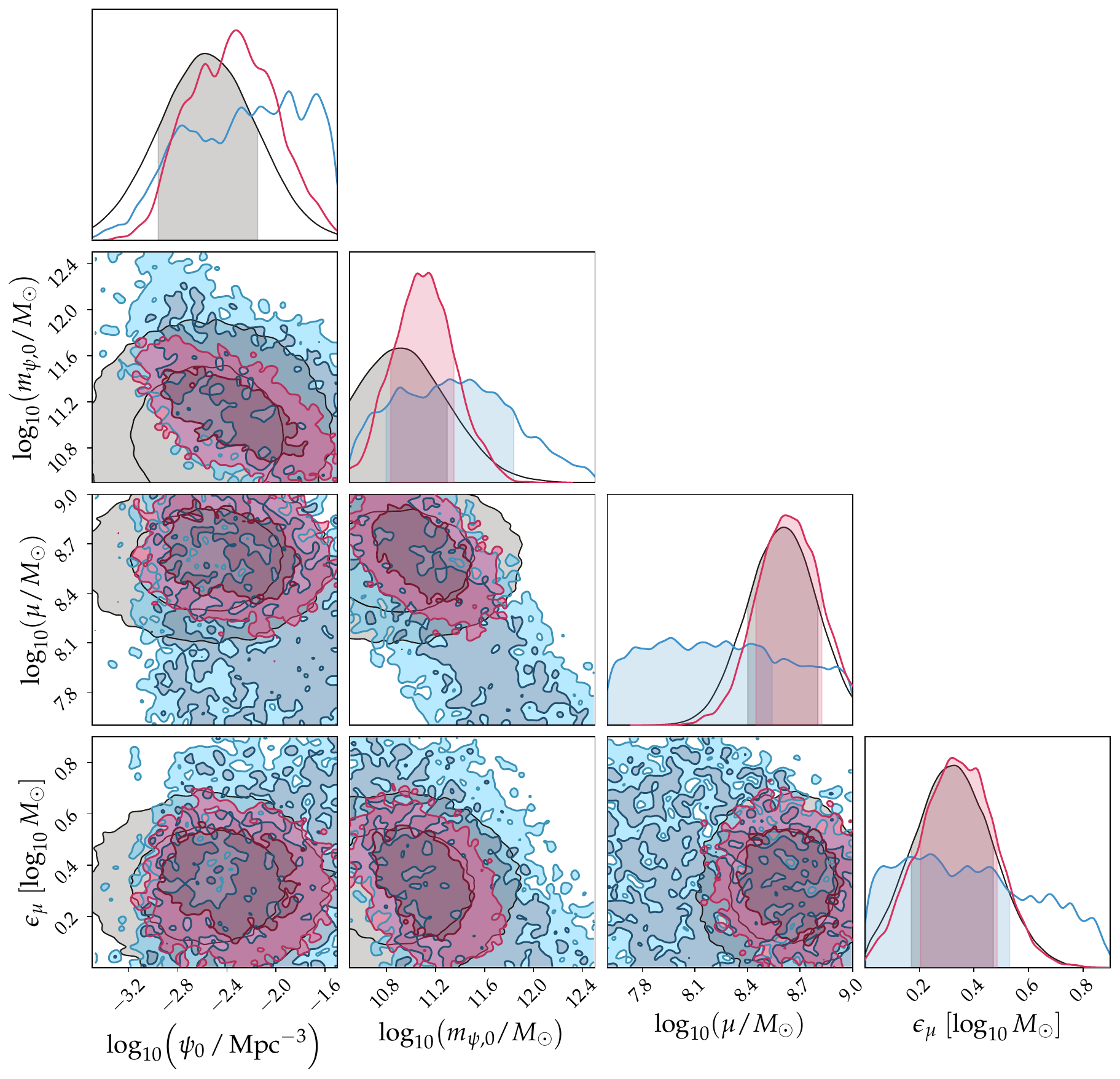}

  \caption{Rejection-resampled posteriors for the GSMF normalization $\psi_0$, GSMF characteristic mass $m_{\psi,0}$, MMB normalization $\mu$ at $M_{\rm bulge}=10^{11}M_\odot$, and intrinsic scatter $\epsilon_\mu$, inferred from the NANOGrav 15-year data. 
  The proposal coordinates $(N_c,h_c)$ are marginalized in obtaining these population-parameter posteriors and therefore do not appear as additional axes.
  Gray contours show results obtained with uniform priors. Blue contours represent astrophysical prior, while pink countours are posteriors with astrophysical priors.
  The original NANOGrav result obtained based on GP interpolation is shown in Figure~9 in Ref.~\cite{NG15_HOLODECK}.
  An agreement with this result supports that SMBHB population constraints can be recovered from the two fundamental degrees of freedom of the GWB in PTA data represented by $(N_c,h_c)$.
  These parameters define our proposal posterior.
  }
  \label{fig:posterior_astro}
\end{figure*}

We present the results of parameter estimation on the NANOGrav 15-year data using the posterior on $(N_c,h_c)$ from Ref.~\cite{GoncharovSato-Polito2026} as a proposal distribution. 
This lets us draw insights from the data using the results of the previous analysis, without processing NANOGrav's pulsar pulse arrival times as part of this work. 
More precisely, the proposal samples indexed by $i$ in Equation~\ref{eq:posterior_reweighted} are obtained from Ref.~\cite{GoncharovSato-Polito2026}, so the posterior represented by this equation contains no data terms.  
The proposal distribution~\cite{GoncharovSato-Polito2026} is based on fitting $\pi(h_{\rm t}|N_{\rm c},h_{\rm c})$ across 15 Fourier frequencies to data, assuming Hellings-Downs correlations of the GWB. 
Based on the methodology described in Subsection~\ref{sec:methods}, we recast posterior samples of $(N_{\rm c},h_{\rm c})$ as $\LamH=\{\psi_0,m_{\psi,0},\mu,\epsilon_\mu\}$ of the semi-analytic SMBHB population model based on the GSMF and the MMB relation.
Figure~\ref{fig:posterior_astro} shows the resulting posteriors for both the uniform and astrophysically motivated priors defined in Table B1 in Ref.~\cite{NG15_HOLODECK}.
For the uniform-prior analysis, the posterior medians and 68\% credible intervals are $\log_{10}(\psi_0/\mathrm{Mpc}^{-3})=-2.22^{+0.49}_{-0.57}$, $\log_{10}(m_{\psi,0}/M_\odot)=11.34^{+0.50}_{-0.49}$, $\log_{10}(\mu/M_\odot)=8.25^{+0.47}_{-0.43}$, and $\epsilon_\mu=0.37^{+0.30}_{-0.24}\,\mathrm{dex}$.
With astrophysically motivated priors, we obtain $\log_{10}(\psi_0/\mathrm{Mpc}^{-3})=-2.35^{+0.35}_{-0.38}$, $\log_{10}(m_{\psi,0}/M_\odot)=11.10^{+0.25}_{-0.25}$, $\log_{10}(\mu/M_\odot)=8.64^{+0.18}_{-0.18}$, and $\epsilon_\mu=0.34^{+0.14}_{-0.14}\,\mathrm{dex}$.
These quantiles computed from normalized weights agree closely with quantiles computed from the finite rejection-resampled sets shown in Figure~\ref{fig:posterior_astro}.
The effective sample size is large in absolute terms and reported in Table~\ref{tab:importance-diagnostics}.

\begin{table}[t]
\centering
\caption{The details from reweighting the posterior on $(N_{\rm c},h_{\rm c})$ of \citet{GoncharovSardana2025a} to $\Lambda_{\rm GSMF}$.}
\label{tab:importance-diagnostics}
\begin{tabular}{lrrrr}
\toprule
Prior & Proposals & $n_{\rm eff}$ & $n_{\rm eff}/n$ & Resampled \\
\midrule
Uniform & $512{,}000$ & $38{,}234$ & $0.0747$ & $3{,}180$ \\
Astrophysical & $406{,}016$ & $54{,}340$ & $0.1338$ & $3{,}751$ \\
\bottomrule
\end{tabular}
\end{table}

The four displayed parameters are those varied in the ``GW-only'' model of Ref.~\cite{NG15_HOLODECK} by NANOGrav, so Figure~\ref{fig:posterior_astro} can be compared directly with the corresponding four population-parameter projections in Figure~9 of Ref.~\cite{NG15_HOLODECK}.
The location, width, and principal degeneracies of our posteriors are consistent with the NANOGrav result, including the trade-off between the GSMF characteristic mass and the normalization of the black-hole--bulge-mass relation.
Unlike NANOGrav, we use a generative model for the strain PDF without interpolating a finite library of Monte Carlo spectra with Gaussian processes.
\citet{Ali-Haimoud2026} further showed that the log-normal approximation used to represent the strain PDF in Ref.~\cite{NG15_HOLODECK} is substantially less accurate than the analytic large-source-count strain PDF.
This issue is distinct from, and additional to, uncertainty from GP interpolation.
The consistency found here indicates that these approximations did not significantly affect the broad population constraints in the four-parameter comparison.
However, our route that avoids both approximations remains more robust.

\begin{figure*}[!htb]
    \centering
    \begin{subfigure}{\textwidth}
    \includegraphics[width=\textwidth]{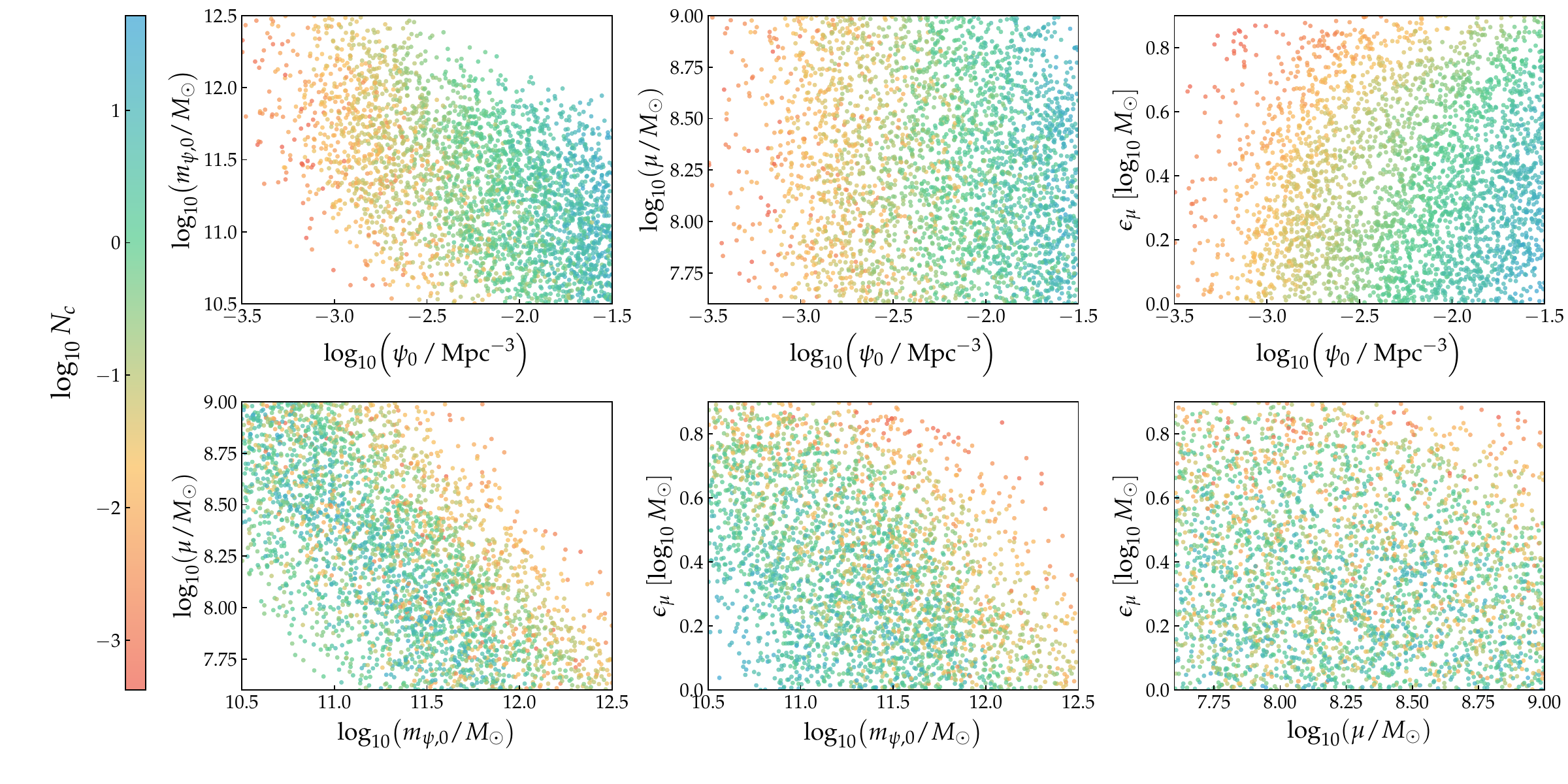}
    \caption{Posterior samples and $N_{\rm c}$}
    \label{fig:population_pairwise_logNc}
  \end{subfigure}
  \hfill
  \begin{subfigure}{\textwidth}
    \includegraphics[width=\textwidth]{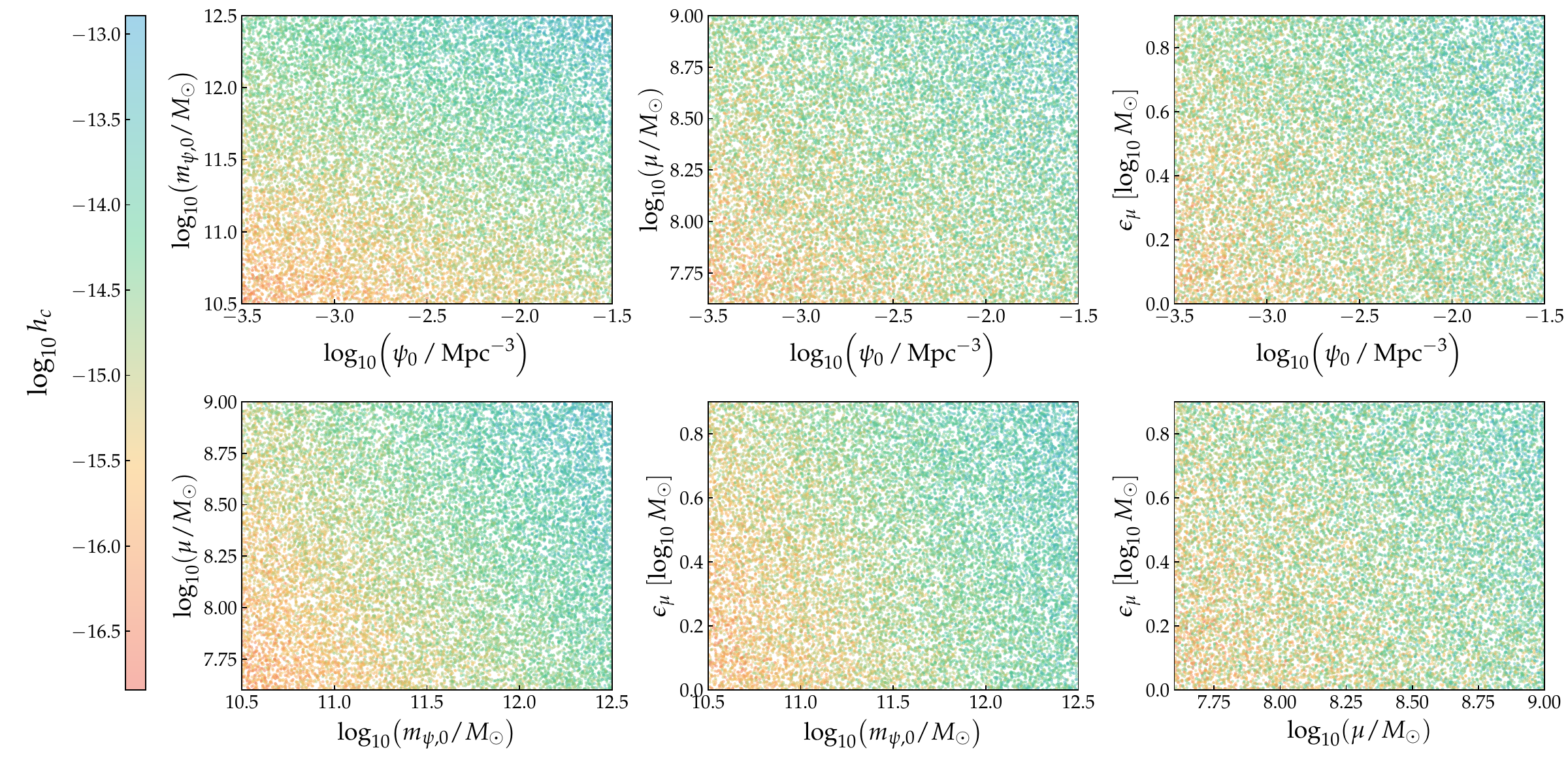}
    \caption{Prior samples and $h_{\rm c}$}
    \label{fig:population_pairwise_loghc}
  \end{subfigure}
    \caption{
    Mapping effective degrees of freedom of the GWB in the PTA band, $(N_{\rm c},h_{\rm c})$ at ${\rm yr}^{-1}$, to SMBHB GSMF-MMB population parameters. 
    \textbf{Top panel}: $N_{\rm c}$ shown for posterior samples from the analysis of the NANOGrav 15-year data.
    The six panels correspond to pairwise projections of the four-dimensional parameter space $(\psi_0,m_{\psi,0},\mu,\epsilon_\mu)$. 
    Here, $\psi_0$ is the GSMF normalization, $m_{\psi,0}$ is the GSMF characteristic mass, $\mu$ is the black-hole--bulge-mass normalization at $M_{\rm bulge}=10^{11}M_\odot$, and $\epsilon_\mu$ is the intrinsic MMB scatter.
    All panels use the same color normalization over the complete sampled range $-3.376\leq\log_{10}N_c\leq1.714$, progressing from coral red through yellow and green to blue as the source count increases.
    The strong ordering with $\psi_0$ shows that the number-density normalization is the principal control on the characteristic source count.
    The remaining panels display how mass and scatter degeneracies redistribute the signal among binaries.
    \textbf{Bottom panel:} $h_{\rm c}$ shown for uniform prior samples from the analysis of the NANOGrav 15-year data, which are used as proposals in rejection resampling. 
    The common colorbar spans the complete proposal range $-16.843\leq\log_{10}h_c\leq-12.892$.
    }
    \label{fig:population_pairwise}
\end{figure*}

Let us examine the connection between the effective GWB parameters $(N_{\rm c},h_{\rm c})$ and GSMF-MMB parameters.
We illustrate this connection in Figure~\ref{fig:population_pairwise}. 
In Figure~\ref{fig:population_pairwise_logNc}, we show two-dimensional marginal posteriors from Figure~\ref{fig:posterior_astro}, where every posterior sample is colored according to its respective $N_{\rm c}$ at ${\rm yr}^{-1}$. 
The figure shows that the dominant variation in $ N_{\rm c} $ follows the GSMF normalization: a larger galaxy number density generally produces a larger characteristic number of contributing binaries.
The characteristic number of sources also varies along combinations of the characteristic galaxy mass and intrinsic scatter, while its dependence on the black-hole--bulge-mass normalization is less nearly monotonic in the pairwise projections.
The color gradients therefore expose which directions within the broad population-parameter degeneracies change the discreteness of the GWB, and what improvements in SMBHB population inference we expect when $N_{\rm c}$ is constrained to a certain range. 
In particular, if we better constrain the GWB discreteness parameter $N_{\rm c}$, the elongated area between $m_{\psi,0}$ and $\psi_0$ in Figure~\ref{fig:population_pairwise_logNc} turns into a more vertical, smaller region. 

Consistently with the results of~\cite{GoncharovSato-Polito2026}, namely, an EPTA\footnote{The European Pulsar Timing Array (EPTA)~\cite{EPTA_DR2_TIMING}.} contour in Figure 8 therein, we find that astrophysical models predict $N_{\rm c}$ to be about $1$ at ${\rm yr^{-1}}$. 
Compared to the original prior $\pi(\log_{10}N_{\rm c})=\mathcal{U}(-3,3)$ of~\citet{GoncharovSato-Polito2026}, the plotted range $-3.376\leq\log_{10}N_c\leq1.714$ in Figure~\ref{fig:population_pairwise_logNc} suggests that semi-analytical SMBHB models do not predict to have as many as $10^3$ characteristic sources at ${\rm yr}^{-1}$. 
More precisely, we find $\log_{10}N_{\rm c}=-0.43 \pm 0.74$ based on the astrophysical GSMF prior, corresponding to the dark blue contour in Figure~\ref{fig:posterior_astro}. 
As shown in Figure~\ref{fig:illustris}, the NANOGrav data excludes $N_{\rm c}$ values significantly lower than $1$ at ${\rm yr}^{-1}$, but remaining consistent with $1$.

Coloring posterior samples according to $h_{\rm c}$ does not show visible gradients. 
The data constrain this amplitude to a comparatively narrow range across the broad allowed population parameter combinations, and high- and low-strain samples consequently remain interspersed in most pairwise projections.
This behavior suggests that different combinations of galaxy abundance, characteristic mass, black-hole mass normalization, and intrinsic scatter can reproduce a similar integrated GWB amplitude.
To demonstrate more clearly how $h_{\rm c}$ affects SMBHB population parameters, we demonstrate $h_{\rm c}$ for prior samples for SMBHB population parameters rather than posterior samples in Figure~\ref{fig:population_pairwise_loghc}. 
The samples fill the four-dimensional parameter domain and recover the expected strain gradients, including increasing $h_c$ toward jointly larger $\psi_0$ and $m_{\psi,0}$.
Together, the two colorings separate the roles of the effective coordinates: $h_c^2$ fixes the integrated first moment of the source-count intensity, whereas $N_c$ describes whether that strain is assembled from many weaker sources or fewer stronger sources.
The explicit gradients across the population posterior make these otherwise compressed parameter-estimation degeneracies directly visible.

Our study focuses on the circular, GW-driven SMBHB population model in which $h_c(f)\propto f^{-2/3}$ in the ensemble mean.
Purely GW-driven population models have a related self-consistency limit when the population is constructed from an SMBHB merger rate: binaries contributing to the GWB must evolve through the PTA band and merge within the available cosmic time. 
\citet{Blecha2026} finds that this approximation remains adequate for current PTA data provided precision at the $\sim1\%$ level is not required, but that the limitation becomes increasingly relevant for longer timing baselines.
The construction of Refs.~\cite{Sato-PolitoZaldarriaga2025b,GoncharovSato-Polito2026} does not yet describe the environmental spectral turnover allowed by the ``phenomenological'' model of Ref.~\cite{NG15_HOLODECK}.
Consequently, the binary-lifetime and phenomenological hardening parameters shown in Figure~9 of that work are neither varied nor constrained here.

\section{\label{sec:conclusions}Conclusions}

We have shown that the four-dimensional population posterior inferred from the NANOGrav 15-year data can be recovered from the two effective parameters $(N_c,h_c)$ introduced by~\citet{Sato-PolitoZaldarriaga2025b} and fitted directly to the PTA data by~\citet{GoncharovSato-Polito2026}.
The recovered constraints agree with the corresponding NANOGrav analysis~\cite{NG15_HOLODECK}, while requiring neither a Gaussian-process interpolation across a population-synthesis library nor an approximation to the model of $\pi(h_{\rm t}|N_{\rm c},h_{\rm c})$. 
This agreement supports the premise that for circular, GW-driven SMBHB populations, the dominant observable variation is described by the mean strain scale $h_{\rm c}$ and the characteristic source count $N_{\rm c}$.
Our calculation uses these quantities as the physical proposal coordinates but evaluates the full population-specific strain PDF, thereby retaining the smaller residual dependence of the rescaled strain-number-density shape.

The parameter space compression changes how high-dimensional population inference can be organized.
The PTA likelihood needs to be explored only once in the two-dimensional $(N_c,h_c)$ space, after which any population parameterization that predicts the source-count intensity can be tested by reweighting the same posterior, provided that the proposal has adequate support.
The computational cost therefore does not inherit the exponential growth of a population-synthesis grid with every additional astrophysical parameter, alleviating the usual curse of dimensionality at the likelihood-evaluation stage.
Additional population parameters may remain weakly identified or exactly degenerate, but their joint posterior structure is exposed by the explicit map from the source-count intensity to $(N_c,h_c)$ rather than absorbed into an interpolator.
This makes it possible to determine which combinations of galaxy abundance, characteristic mass, black-hole scaling, and intrinsic scatter are actually measured by the PTA data, and which are separated only by external astrophysical priors.

The parameter space reduction is conditional on the assumptions underlying Refs.~\cite{Sato-PolitoZaldarriaga2025b,GoncharovSato-Polito2026}: circular binaries, GW-driven evolution, and the corresponding power-law mean characteristic strain.
Extending the generative strain PDF to environmental hardening, eccentricity, or other mechanisms that produce a spectral turnover is required before the same claim can be made for the ``phenomenological''  model parameters from Ref.~\cite{NG15_HOLODECK}.
Within its current domain, however, our method provides a common and transparent connection between PTA measurements and semi-analytic or simulation-based SMBHB populations: the $(N_c,h_c)$ analysis supplies a broadly supported proposal, and each target model supplies its full strain PDF without requiring a model-specific spectral emulator.

Our calculation of the source-count intensity for Illustris demonstrates the methodology to map the results of cosmological simulations to PTA data. 
With the Illustris catalogue of galaxy mergers provided with \textsc{holodeck}, we find $h_{\rm c}(1\,\mathrm{yr}^{-1})=(3.40$--$4.27)\times10^{-16}$ across the explored lifetimes, approximately a factor of five to six below the median inferred from the NANOGrav 15-year data.
The result provides conditional predictions for circular binaries evolved with the adopted phenomenological hardening prescription at each specified value of $\tau$.
Constructing a robust calibrated Illustris prediction would require a physical probability distribution for the unresolved lifetime and propagation of the finite-catalogue and population-realization uncertainty, with other assumptions such as eccentricity or alternative hardening models treated as separately stated conditional predictions or assigned explicit prior weights.
The spatial block-resampling region presented here propagates the within-box finite-catalogue component at fixed $\tau$, but it does not replace independent cosmological realizations or quantify systems and long-wavelength modes absent from the finite Illustris volume.

Our methodology may be of interest to apply to other numerical hydrodynamic simulations of galaxy mergers, to extend the predictions for $h_{\rm c}(f)$ to additionally $N_{\rm c}(f)$. 
ASTRID includes an on-the-fly dynamical-friction treatment for black holes and predicts $h_c\simeq2.8\times10^{-15}$ near 4 nHz.
\citet{ChenDiMatteo2025} find that omitting this treatment instead produces factors of 3--5 excess merger-driven black-hole growth and a comparable upward bias in the GWB prediction.

Horizon-AGN attributes its dominant signal to hundreds to thousands of binaries with chirp masses $10^{8.5}$--$10^{9.5}M_\odot$ at $z=0.05$--1 and cautions that power-law inference can overstate an amplitude discrepancy \cite{QuelquejayLeclereLi2026}.

In FABLE, the fiducial population differs from PTA posteriors by only $1$--$2.5\sigma$, with the inferred discrepancy shifting by several tenths of a standard deviation under alternative pulsar-noise models \cite{ButtigiegSijacki2026}.

\section*{Acknowledgements}

We thank Siyuan Chen and Laura Blecha for helpful comments. 
We use code \textsc{holodeck} available at \href{https://github.com/nanograv/holodeck}{github.com/nanograv/holodeck} to build the luminosity function for the semi-analytic GSMF approach and to access and evolve the results of the ``Illustris'' simulation.

\bibliography{mybib,collab_papers,soft}{}
\bibliographystyle{apsrev4-2}

\appendix

\end{document}